\documentclass[onecolumn]{article}    

\usepackage{graphicx}          
                               
\usepackage{amssymb}
\usepackage{amsthm}
\usepackage{subfig}
\usepackage{mathtools}
\usepackage{authblk}
\usepackage{hyperref}

\usepackage{bm}
\usepackage{amsmath}
\usepackage{amsfonts}
\usepackage{subfig}

\usepackage{mathabx}
\usepackage{blindtext}

\usepackage{multirow}

\usepackage[a4paper, rmargin = 2.5cm, lmargin = 2.5cm]{geometry}

\renewcommand\vec{\mathbf}

\usepackage{float}
\usepackage{epstopdf}
\AppendGraphicsExtensions{.tif}

\title{A Circular Restricted n-body Problem}

\author{Rodolfo Batista Negri \footnote{Ph.D. Candidate, rodolfo.negri@inpe.br, Av. dos Astronautas 1758, São José dos Campos.} and Antônio F. B. A. Prado \footnote{Postgraduate Division, National Institute for Space Research (INPE), 12227-010, São José dos Campos, SP, Brazil. Volunteer Professor, Academy of Engineering, RUDN University, Miklukho-Maklaya street 6, Moscow, Russia, 117198. Associate Fellow AIAA.}}
\affil{National Institute for Space Research, 12227-010, São José dos Campos, São Paulo, Brazil}

\begin{document}

\maketitle





\section{Introduction}

The use of dynamical approximations to design spacecraft trajectories is a procedure that goes back to the beginnings of the astrodynamics. By the 1920s, Tsander \cite{tsander1964} was the first to use two-body assumptions to propose the use of gravity-assists for interplanetary missions, in what we would call today a patched-conics approximation~\cite{negri2020historical}. The patched-conics approximates a three-body problem by breaking the problem into a set of two-body problems and then patching the separately obtained Keplerian trajectories together to create the whole trajectory. These kinds of approximations would reemerge in the 50s, in the rise of the space age, with the contribution of different astrodynamicists when also proposing ``perturbation maneuvers'' or ``gravitational navigation'' (how they referred to gravity-assists) \cite{lawden1954perturbation,crocco1956one,ehricke1957instrumented,minovitch1961method,negri2020historical}. Its application can describe better or worse the physical phenomenon depending on the system in which it is applied, and other parameters involved in the dynamics, as it is the case in any approximation~\cite{negri2017studying,negri2019lunar}.

Another approximation for a three-body problem present in the beginnings of the space age is the circular restricted three-body problem (CR3BP)~\cite{szebehely2012theory}. The work of Egorov~\cite{egorov1958certain} using the CR3BP was the base to design the trajectory of the spacecraft Luna 3. These early astrodynamicists did not limit themselves to these approximations as new space mission proposals and technical difficulties were assessed. The famous scientist and space enthusiast Krafft Ehricke showed that considering the gravitational pull of the sun in a restricted four-body problem is very important for a cislunar ``instrumented comet'' (interplanetary spacecraft) \cite{ehricke1957instrumented}. In this context of early findings in astrodynamics, Huang \cite{huang1960very} derives an approximation for a four-body problem in order to support that the CR3BP could still be used for designing trajectories in the earth-moon system without considering the gravitational pull of the sun. Huang's equations are known today in the astrodynamics community as the bicircular restricted four-body problem (BCR4BP), in which a system of two massive bodies (called primaries) are assumed to describe a Keplerian motion between themselves about their barycenter. A third massive body is added and considered to follow a two-body problem with the mass of both primaries as if they both were in their barycenter~\cite{negri2020generalizing}. The fourth body is the infinitesimal mass particle immersed in the gravitational field generated by the other three.

The BCR4BP has been extensively and successfully applied to design trajectories as a straightforward to use model since Huang's derivation, primarily applied as a tool for designing trajectories in multiple body systems \cite{howell1986periodic,belbruno1993sun,castella2000vertical,mingotti2011optimal,qi2014gravitational,heiligers2018solar,negri2019lunar,boudad2020dynamics}. While most of these applications are to binary systems\footnote{Here we define a binary system as a system of primaries that are several orders of magnitude closer to one another than to the third body.} of primaries (e.g., earth-moon, binary asteroid), others extended the application to non-binary systems \cite{gabern2001restricted,pergola2009earth,oshima2015jumping,barrabes2016pseudo}. However, the BCR4BP as derived by Huang \cite{huang1960very} is specifically derived for binary systems, holding an incorrect simplification when dealing with non-binary systems of primaries~\cite{negri2020generalizing}. Negri and Prado~\cite{negri2020generalizing} proposed a generalization of the equations by supposing more coherent two-body approximations and showing that the equations reduce to the usual BCR4BP when applied to binaries. 

In this work, we extend the approach taken by Negri and Prado~\cite{negri2020generalizing} to expand the BCR4BP equations to describe an n-body system. Like the BCR4BP, we impose the artificial constraint that each massive body in the system can only describe a Keplerian motion, and two-body problems are approximated accordingly. The Keplerian orbits of the massive bodies are assumed as coplanar and circular. Although the reasoning in expanding the BCR4BP to a CRNBP seems trivial, the only attempt (to our best knowledge) before made in the literature is the one by Iuliano \cite{iuliano2016solution}, which neglect some indirect terms. We show with illustrative examples that the CRNBP seems to retain the complex behavior of an ephemerides n-body problem. Its simplicity and resemblance to a CR3BP make it an excellent choice for designing complex trajectories in multiple body systems, with special applications to the outer planets. A much faster and simpler preliminary design is possible with the CRNBP, needing to integrate only six first-order ordinary differential equations instead of the 6N of an ephemerides model. We choose in this paper to go straight to the point and let extensive analysis on any application to dedicated future works. We also make this straightforward approach hoping that the astrodynamics community can join in the effort of exploring the trajectory design possibilities enabled by the model.

\section{Main Results}

\subsection{Deirvation of the CRNBP}

\begin{figure}
	\centering\includegraphics[width=.5\textwidth]{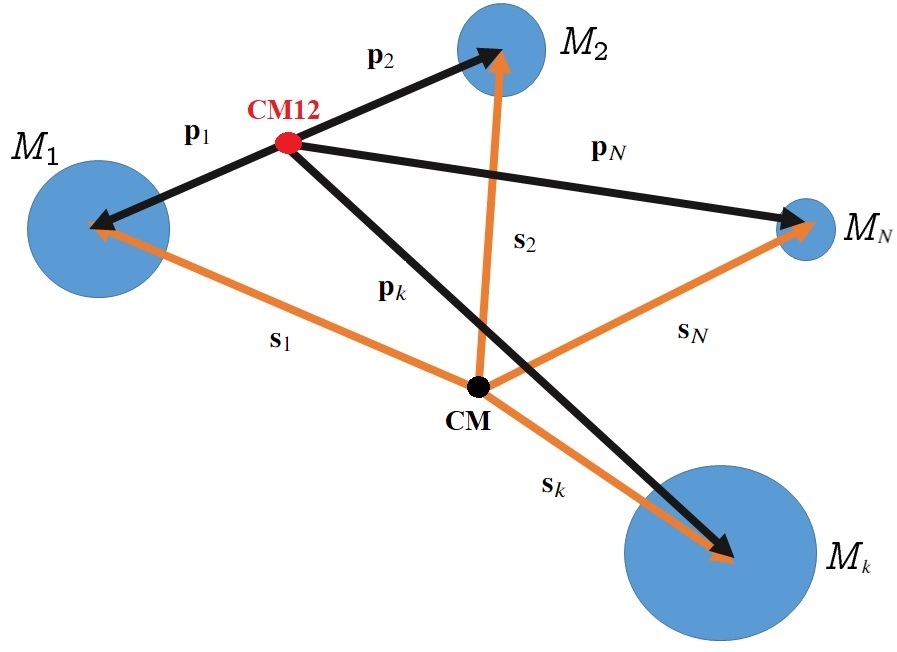}
	\caption{Representation of a n-body Problem.}
	\label{fig:Fig1}
\end{figure}

Assume the gravitational interaction between $N$ bodies of masses $M_j$, for $j=1,2,...,k,...,N$. Consider an inertial reference frame placed in their center of mass (CM) with $\vec{s}_j$ representing the position of each of these bodies in that frame, while vectors $\vec{p}_j$ define their position with respect to the barycenter of the masses $M_1$ and $M_2$ (CM12), as shown in Fig. \ref{fig:Fig1}. Defining the relative position between the masses $M_k$ and $M_j$ as $\vec{s}_{jk} = \vec{s}_k - \vec{s}_j$, we can obtain the equation of motion of each mass $M_j$ as:

\begin{equation}
\ddot{ \vec{ s}} _j = G \sum_{k=1,k\neq j}^{N} M_k \frac{\vec{s}_{jk}}{\vert \vert \vec{s}_{jk} \vert \vert ^3}, \text{\hspace{.2cm}} j=1,2,...,N.
\end{equation}

We now assume a restricted problem, in which the N-th body has a mass much smaller than the others:

\begin{subequations}
\label{eqn:restricted_nbody}
\begin{align}
\label{eq:nbody_massives}
\ddot{\vec{s}}_j &= G \sum_{k=1,k\neq j}^{N-1} M_k \frac{\vec{s}_{jk}}{\vert \vert \vec{s}_{jk} \vert \vert ^3}, \text{\hspace{.2cm}} j=1,2,...,N-1, \\
\label{eq:nbody_restr}
\ddot{\vec{s}}_N &= G \sum_{j=1}^{N-1} M_j \frac{\vec{s}_{Nj}}{\vert \vert \vec{s}_{Nj} \vert \vert ^3}.
\end{align}
\end{subequations}
Noting that $\vec{p}_j=\vec{s}_j-\vec{s}_{CM12}$, Eq. \eqref{eq:nbody_restr} can be rewritten in a reference frame centered in CM12 as:

\begin{equation}
\label{eq:pN_w_sCM12}
\ddot{\vec{p}}_N + \ddot{\vec{s}}_{CM12}  = G \sum_{j=1}^{N-1} M_j \frac{\vec{p}_j-\vec{p}_N}{\vert \vert \vec{p}_j-\vec{p}_N \vert \vert ^3} .
\end{equation}
Since it is assumed that the inertial frame is centered in the barycenter of the n-body system, we have that $\vec{s}_{CM} = M_1 \vec{s}_1 + M_2 \vec{s}_2 + ... + M_N \vec{s}_N = 0$. Therefore, the position of CM12 can be rewritten following the relation:

\begin{equation}
\vec{s}_{CM12} = \frac{1}{M_1 + M_2} ( M_1 \vec{s}_1 + M_2 \vec{s}_2 )  = - \sum_{j=3}^{N-1} \frac{M_j}{M_1 + M_2 } \vec{s}_j ,
\end{equation}
which may be differentiated twice for obtaining its equation of motion:

\begin{equation}
\label{eq:sCM12_1st}
\ddot{\vec{s}}_{CM12} = - \sum_{j=3}^{N-1} \frac{M_j}{M_1 + M_2 }\ddot{ \vec{s}}_j .
\end{equation}
Now, Eq. \eqref{eq:nbody_massives} can be coupled into Eq. \eqref{eq:sCM12_1st}, remembering that $\vec{s}_j=\vec{p}_j+\vec{s}_{CM12}$, to arrive to:

\begin{equation}
\label{eq:eqn_mot_CM12}
\ddot{\vec{s}}_{CM12} = - \frac{G}{M_1 + M_2} \sum_{j=3}^{N-1}  M_j \sum_{k=1,k \neq j}^{N-1} M_k \frac{\vec{p}_k-\vec{p}_j}{\vert \vert \vec{p}_k-\vec{p}_j \vert \vert ^3} .
\end{equation}
Finally, Eq. \eqref{eq:eqn_mot_CM12} can be used for writing Eq. \eqref{eq:pN_w_sCM12} with terms relative only to the CM12:

\begin{equation}
\ddot{\vec{p}}_N   = G \sum_{j=1}^{N-1} M_j \frac{\vec{p}_j-\vec{p}_N}{\vert \vert \vec{p}_j-\vec{p}_N \vert \vert ^3} + \frac{G}{M_1 + M_2} \sum_{j=3}^{N-1}  M_j \sum_{k=1,k \neq j}^{N-1} M_k \frac{\vec{p}_k-\vec{p}_j}{\vert \vert \vec{p}_k-\vec{p}_j \vert \vert ^3} ,
\end{equation}
which can be easily rearranged as:

\begin{align}
\label{eqn:p_N_fim}
\begin{split}
\ddot{\vec{p}}_N   &= G \left[ M_1 \frac{\vec{p}_1 - \vec{p}_N}{\vert\vert\vec{p}_1 - \vec{p}_N\vert\vert^3} + M_2 \frac{\vec{p}_2 - \vec{p}_N}{\vert\vert\vec{p}_2 - \vec{p}_N\vert\vert^3} \right. \\ & \left. +  \sum_{j=3}^{N-1} M_j \left(  \frac{\vec{p}_j - \vec{p}_N}{\vert\vert\vec{p}_j - \vec{p}_N\vert\vert^3}  +  \sum_{k=1,k\neq j}^{N-1} \frac{M_k}{M_1 + M_2} \frac{\vec{p}_k - \vec{p}_j}{\vert\vert\vec{p}_k - \vec{p}_j\vert\vert^3} \right) \right] .
\end{split}
\end{align}

For now, the only approximation made was to assume an infinitesimal mass and neglect its gravitational influence on the other bodies. However, inspired by the BCR4BP, we can impose other constraints on the system to simplify its description, analysis, and application. Negri and Prado~\cite{negri2020generalizing} framework for generalizing the BCR4BP also allows to extend the equations to n-bodies, and that is what we make now. 

\begin{figure}
	\centering\includegraphics[width=.5\textwidth]{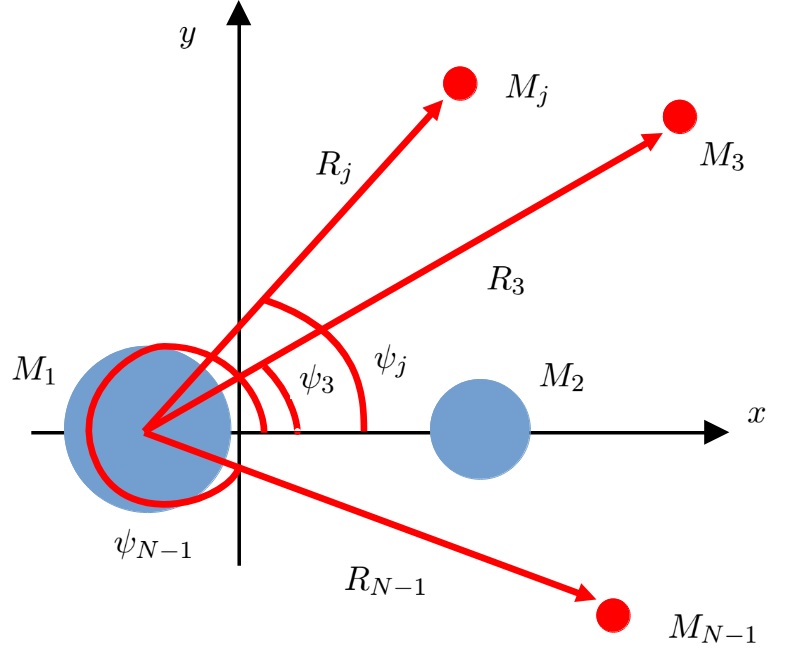}
	\caption{Representation of the Circular Restricted n-body Problem, in the synodic frame.}
	\label{fig:Fig2}
\end{figure}

First, it is assumed that the main primaries, defined as the bodies of mass $M_1$ and $M_2$, describe a circular orbit around their common center of mass (CM12). The remainder massive bodies are assumed to describe circular orbits centered in the body $M_1$. An additional constraint is added by assuming that all the orbits of the massive bodies are coplanar. Now, assuming that the system centered in CM12 is synodic with an angular velocity equal to the one of $M_1$ and $M_2$, Eq. \eqref{eqn:p_N_fim} becomes:

\begin{align}
\label{eqn:rho_synodic}
\begin{split}
\ddot{\vec{\rho}}_N   &= - 2 \vec{\Omega} \times \dot{\vec{\rho}} - \vec{\Omega} \times (\vec{\Omega} \times \vec{\rho}) + G \left[ M_1 \frac{\vec{\rho}_1 - \vec{\rho}_N}{\vert\vert\vec{\rho}_1 - \vec{\rho}_N\vert\vert^3} + M_2 \frac{\vec{\rho}_2 - \vec{\rho}_N}{\vert\vert\vec{\rho}_2 - \vec{\rho}_N\vert\vert^3} \right. \\ & \left. +  \sum_{j=3}^{N-1} M_j \left(  \frac{\vec{\rho}_j - \vec{\rho}_N}{\vert\vert\vec{\rho}_j - \vec{\rho}_N\vert\vert^3}  +  \sum_{k=1,k\neq j}^{N-1} \frac{M_k}{M_1 + M_2} \frac{\vec{\rho}_k - \vec{\rho}_j}{\vert\vert\vec{\rho}_k - \vec{\rho}_j\vert\vert^3} \right) \right] ,
\end{split}
\end{align}
where $\vec{\rho}_N = \begin{bmatrix} x & y & z \end{bmatrix}^\text{T}$ represents the position of the infinitesimal mass body in the synodic frame, and $\vec{\Omega}$ is the angular velocity of the synodic frame relative to the fixed one. The degrees of freedom (DOF) of $M_1$ and $M_2$ can be eliminated with these approximations by noting they remain fixed in the synodic frame, while the other massive bodies are reduced to a single DOF each, as shown in Fig. \ref{fig:Fig2}. Assuming the same canonical units used in the BCR4BP and CR4BP~\footnote{The units of mass, distance and time are normalized such that: the mass parameter $\mu_j = M_j / (M_1+M_2)$ is the canonical mass of $M_j$; the distance between $M_1$ and $M_2$ is unitary; and the angular speed of the synodic frame is $n_{12}=1$.}, it is easy to find that the distance from each massive body to the particle and between any two massive bodies are, respectively:

\begin{subequations}
\label{eqn:rho_positions}
\begin{align}
\vec{\rho}_j - \vec{\rho}_N &= \begin{bmatrix}
R_j cos \psi_j - \mu_2 -x \\ R_j\sin \psi_j - y \\ -z
\end{bmatrix}, \\
\vec{\rho}_k - \vec{\rho}_j &= \begin{bmatrix}
R_k \cos \psi_k-R_j \cos\psi_j \\
R_k \sin \psi_k-R_j \sin\psi_j \\
0
\end{bmatrix}.
\end{align}
\end{subequations}
in which $R_j$ represents the distance between each massive body and $M_1$, and $\psi_j$ is a phase angle defined from x-axis of the synodic frame, as shown in Fig. \ref{fig:Fig2}, for $j = 1,2,...,N-1$, $k \neq j$. Note that in the synodic frame $R_1=0$, $R_2=1$, and $\psi_2 = 0 \degree$.

Finally, substituting Eqs. \eqref{eqn:rho_positions} into Eq. \eqref{eqn:rho_synodic} and using elementary algebra, one can find the equations of motion for the CRNBP:

\begin{subequations}
\label{eq:N2CRNBP}
\begin{align}
\begin{split}
\ddot{x} &= 2 \dot{y} + x - \frac{\mu_1}{r_1^3}(x+\mu_2) - \frac{\mu_2}{r_2^3} (x-\mu_1) - \sum_{j=3}^{N-1} \mu_j \left[ \frac{1}{r_j^3} (x+\mu_2-R_j \cos \psi_j ) + \right. \\ & \left.  \sum_{k=1,k\neq j}^{N-1} \frac{\mu_k}{(R_k^2+R_j^2-2 R_k R_j \cos(\psi_k-\psi_j) )^{3/2}} (R_j \cos \psi_j - R_k \cos \psi_k ) \right], 
\end{split} \\
\begin{split}
\ddot{y} &= - 2 \dot{x} + y - \frac{\mu_1}{r_1^3} y - \frac{\mu_2}{r_2^3} y - \sum_{j=3}^{N-1} \mu_j \left[ \frac{1}{r_j^3} (y-R_j \sin \psi_j ) + \right. \\ & \left.  \sum_{k=1,k\neq j}^{N-1} \frac{\mu_k}{(R_k^2+R_j^2-2 R_k R_j \cos(\psi_k-\psi_j) )^{3/2}} (R_j \sin \psi_j - R_k \sin \psi_k ) \right], 
\end{split} \\
\ddot{z} &= - \frac{\mu_1}{r_1^3} z - \frac{\mu_2}{r_2^3} z - \sum_{j=3}^{N-1}  \frac{\mu_j}{r_j^3} z. 
\end{align}
\end{subequations}
$r_j$, $j=1,2,...,N-1$, is the magnitude of the respective $\vec{\rho}_j-\vec{\rho}_N$, and the mass parameter of each body is defined as $\mu_j=M_j/(M_1+M_2)$. The $\psi_j$ can be solved in time by the analytical expression:
\begin{equation}
\label{eq:psi_j}
\psi_j = \psi_{0j}+ (n_j-n_{12}) t,
\end{equation}
where $\psi_{0j}$ represents the initial phase angle, $n_j$ is the mean motion of the $j$-th body in canonical units (if the motion is retrograde relative to the primaries, it must be negative), and $n_{12}$ is the mean motion of $M_1$ and $M_2$ about their center of mass, which is $n_{12}=1$ in canonical units.

Note that this problem description reduces the 6N first-order differential equations of a restricted n-body problem to simply six differential equations and N-2 analytical expressions. Moreover, the problem is cast in a framework similar to the CR3BP, making it an excellent intermediate model between a CR3BP and a general ephemerides model in systems of multiple bodies (e.g., outer planetary systems). The difference between Eqs. \eqref{eq:N2CRNBP} and the equations derived by Iuliano \cite{iuliano2016solution} is in the summation inside the brackets. That summation is part of the indirect gravitational effect of each massive body pulling one another. That term is important to correctly approximate the indirect effect, as shown in Negri and Prado \cite{negri2020generalizing} for the generalization of the BCR4BP. Another important fact is that Eqs. \ref{eq:N2CRNBP} reduce to the ``conventional'' BCR4BP, referred in Negri and Prado \cite{negri2020generalizing} as the ``binary case'', if $N=4$ and $R_3 \gg 1$.



\subsection{Ephemerides Correspondence}
\label{sec:ephemerides}

The CRNBP could be an excellent intermediary step for designing complex trajectories before committing to an ephemerides n-body problem. However, it remains the question of how to relate the simplified CRNBP with the ephemerides of the intended application. Therefore, a proposal is presented to relate the ephemerides with the CRNBP variables. First, consider the average orbital elements of the mass body $M_2$ with respect to a fixed reference frame centered in $M_1$, and define the following unit vectors:

\begin{subequations}
\label{eq:iygyhuh}
\begin{align}
    \hat{h}_2 &= \begin{bmatrix}\sin \Bar{i}_2 \sin \Bar{\Omega}_2 \\
    -\sin \Bar{i}_2 \cos \Bar{\Omega}_2 \\
    \cos \Bar{i}_2 \end{bmatrix},\\
    \hat{e}_2 &= \begin{bmatrix}\cos \Bar{\omega}_2 \cos \Bar{\Omega}_2 - \sin \Bar{\omega}_2 \sin \Bar{\Omega}_2 \cos \Bar{i}_2 \\
    \cos \Bar{\omega}_2 \sin \Bar{\Omega}_2 + \sin \Bar{\omega}_2 \cos \Bar{\Omega}_2 \cos \Bar{i}_2 \\
    \sin \Bar{\omega}_2\sin \Bar{i}_2 \end{bmatrix}, \\
    \hat{e}_{\perp 2} &= \hat{h}_2 \times \hat{e}_2
\end{align}
\end{subequations}
where $\hat{h}_2$ is the angular momentum unit vector of $M_2$, and $\hat{e}_2$ is its Runge-Lenz-Laplace unit vector. The mean inclination, argument of periapsis and longitude of the ascending node of $M_2$ in the chosen fixed frame are, respectively: $\Bar{i}_2$, $\Bar{\omega}_2$ and $\Bar{\Omega} _2$.

Given an ephemerides configuration in the fixed system, the initial positions for each massive body $M_j$, $j=1,2,...,N-1$, can be projected onto the orbital plane of $M_2$ according to :
\begin{equation}
\text{proj}_{\mathcal{O}_2} \vec{s}_{1j} = (\vec{s}_{1j} \cdot \hat{e}_2) \hat{e}_2 + (\vec{s}_{1j} \cdot \hat{e}_{\perp 2}) \hat{e}_{\perp 2},
\end{equation}
$\mathcal{O}_2=\left\{ \vec{s} \in \mathbb{R}^3 \vert \vec{s} \cdot \hat{h}_2 =0 \right\}$. 

It is common to obtain ephemeris data about the most massive body. In this case, by the definitions already adopted, $M_1$. Due to that reason, the notation of the vectors of the fixed frame chosen for the ephemerides is given relative to the body of mass $M_1$, $\vec{s}_{1j}$~\footnote{If a different procedure is adopted, other considerations must be made and performed in some of the steps presented.}. It is now possible to establish the initial configuration of each massive body as:

\begin{equation}
\psi_{j0} = \arctan \left[ \frac{(\text{proj}_{\mathcal{O}_2} \vec{s}_{12} \times \text{proj}_{\mathcal{ O}_2} \vec{s}_{1j})\cdot \hat{z}}{(\hat{h}_{2}\cdot \hat{z})({proj}_{\mathcal{ O}_2} \vec{s}_{12} \cdot \text{proj}_{\mathcal{O}_2} \vec{s}_{1j})} \right].
\end{equation}
The mean motion of each body, $n_j$, used in Eq. \eqref{eq:psi_j} is calculated from the bodies' mean orbital period in the fixed system.

For simplicity, it is assumed that the correspondence in ephemerides of the massive bodies to be at the time $t=0$ used for Eqs. \eqref{eq:N2CRNBP}. Now, consider a set of unit vectors defining a second fixed frame that is coincident with the rotating frame at the instant $t=0$, given as follows:

\begin{subequations}
\label{eq:CRNBP_sist_gir_unit vector}
\begin{align}
\hat{S}_1 = & \frac{\text{proj}_{\mathcal{O}_2} \vec{s}_{12} }{\vert \vert \text{proj}_{\mathcal{ O}_2} \vec{s}_{12} \vert \vert}, \\
\hat{S}_2 = & \hat{h}_2 \times \hat{S}_1, \\
\hat{S}_3 = & \hat{h}_2.
\end{align}
\end{subequations}
Then, if necessary, it is possible to transform from the fixed frame adopted for the ephemerides to that second fixed frame coincident with the rotating one in $t=0$, which can be done by using the following rotation matrix:
\begin{equation}
S = \begin{bmatrix}
\hat{S}_1^\text{T} \\
\hat{S}_2^\text{T} \\
\hat{S}_3^\text{T}
\end{bmatrix}.
\end{equation}

As the synodic frame rotates about its axis $z$, it is simple to transform the rotating system from time $t=0$ to time $t=t_N$, just by applying the rotation matrix:
\begin{equation}
\label{eq:CRNBP_T}
T =
\begin{bmatrix}
\cos (n_{12} t_N) & \sin (n_{12} t_N) & 0 \\
-\sin (n_{12} t_N) & \cos (n_{12} t_N) & 0 \\
0 & 0 & 1
\end{bmatrix}.
\end{equation}
Note that $n_{12}=1$ when using canonical units, which may be disregarded in later steps for that case.

If there is also the need to obtain an ephemeris correspondence for the infinitesimal mass body, it is possible to transform its position and velocity from the fixed frame used for the ephemerides to the rotating one. Therefore, with the vectors $\vec{s}_{1N}$ and $\dot{\vec{s}}_{1N}$ obtained from an ephemerides table, it can be shown that:

\begin{subequations}
\label{eq:CRNBP_s1N_to_pN}
\begin{align}
\vec{p}_N = & \vec{s}_{1N} - \mu_2 \vec{s}_{12}, \\
\dot{\vec{p}}_N = & \dot{\vec{s}}_{1N} - \mu_2 \dot{\vec{s}}_{12}.
\end{align}
\end{subequations}
The vectors $\vec{s}_{12}$ and $\dot{\vec{s}}_{12}$ must be obtained considering the approximation of circular and coplanar orbits to maintain consistency with the CRNBP simplifications. For that matter, first consider transforming the vectors $\vec{p}_j$ and $\dot{\vec{p}}_j$ to the synodic frame:
\begin{subequations}
\label{eq:CRNBP_transf_p_to_rho}
\begin{align}
\vec{\rho}_j = & T S \vec{p}_j , \\
\dot{\vec{\rho}}_j = & \dot{T} S \vec{p}_j + T S \dot{\vec{p}}_j.
\end{align}
\end{subequations}
Note that, in the synodic frame, the bodies $M_1$ and $M_2$ remain at rest, and their positions are, respectively, $\vec{\rho}_1=-\mu_2 \hat{x}$ and $\vec{\rho}_1=\mu_1 \hat{x}$, in canonical units. The $\hat{x}$ represents the unit vector defining the $x$-coordinate in the synodic frame, i.e., $\hat{x}=\begin{bmatrix} 1 & 0 & 0  \end{bmatrix}^\text{T}$. Now, one can use the Eqs. \eqref{eq:CRNBP_transf_p_to_rho} to find that:
\begin{subequations}
\begin{align}
\vec{p}_1= &-\mu_2 S^\text{T} T^\text{T} \hat{x}, \\
\vec{p}_2= &\mu_1 S^\text{T} T^\text{T} \hat{x}, \\
\dot{\vec{p}}_1= &\mu_2 S^\text{T} T^\text{T} \dot{T} T^\text{T} \hat{x}, \\
\dot{\vec{p}}_2= &-\mu_1 S^\text{T} T^\text{T} \dot{T} T^\text{T} \hat{x}.
\end{align}
\end{subequations}
using the fact that, for an arbitrary rotation matrix $R$, its inverse is equal to its transpose, $R^{-1}=R^\text{T}$.

Now, one can approximate $\vec{s}_{12}$ and $\dot{\vec{s}}_{12}$, in order to maintain consistency with the CRNBP, according to:
\begin{subequations}
\label{eq:CRNBP_rho12_to_s12}
\begin{align}
\vec{s}_{12}=& \vec{p}_2 - \vec{p}_1 = S^\text{T} T^\text{T} \hat{x}, \\
\dot{\vec{s}}_{12}=& \dot{\vec{p}}_2 - \dot{\vec{p}}_1 = - S^\text{T} T^\text{ T} \dot{T} T^\text{T} \hat{x}.
\end{align}
\end{subequations}
Therefore, it is possible to take from the fixed ephemerides to the synodic frame, through the application of Eqs. \eqref{eq:CRNBP_s1N_to_pN}, \eqref{eq:CRNBP_transf_p_to_rho}, and \eqref{eq:CRNBP_rho12_to_s12}, as described by the following expression:
\begin{subequations}
\begin{align}
\vec{\rho}_N = & T S \left( \vec{s}_{1N} - \mu_2 S^\text{T} T^\text{T} \hat{x} \right), \\
\dot{\vec{\rho}}_N = & \dot{T} S \left( \vec{s}_{1N} - \mu_2 S^\text{T} T^\text{T} \hat{x} \right) + T S \left( \dot{\vec{s}}_{1N} + \mu_2 S^\text{T} T^\text{ T} \dot{T} T^\text{T} \hat{x} \right).
\end{align}
\end{subequations}

Finally, considering that the ephemeris of the infinitesimal body might be obtained at an epoch $JD$, different from the one used for the ephemerides of the massive bodies, one can find $t_N$ in canonical units as:
\begin{equation}
t_N = 86400(JD - JD_0) n_{12},
\end{equation}
where $n_{12}$ is the mean motion of $M_2$, but now in rad/s, as its role here is to transform from seconds to units of canonical time, and $JD_0$ is the Julian date of the ephemerides of the massive bodies. Note that such a transformation is only valid assuming $t=0$ as the time when the ephemerides of the massive bodies are obtained. The initial time for the integration of Eqs. \eqref{eq:N2CRNBP}, when considering the ephemeris of the infinitesimal body, is $t_N$. We make that distinction because it might be helpful in some applications to consider the ephemeris of the infinitesimal body in a different epoch than the one used for the massive bodies.

\section{Illustrative Examples}

As discussed by Wiggins \cite{wiggins1990introduction}, the structural stability of a dynamical system (whether mathematically close dynamical systems have qualitatively the same dynamics) for the applied scientist is still an open question, varying from case to case. Due to this challenge, it is difficult to definitively establish the structural stability of the CRNBP to an ephemerides n-body problem. We then assess its structural stability by reproducing a complex dynamical behavior observed in an ephemerides n-body model system. For that sake, we apply the CRNBP and try to reproduce the results obtained by Todorovic et al. \cite{todorovic2020arches}, in which a chaos indicator, the fast Lyapunov indicator ($FLI$) \cite{froeschle1997fast}, is applied to the solar system using an n-body problem with ephemerides from September 30, 2012, 00:00:00 TDB, as the initial condition. For the CRNBP, we consider the same bodies used in Todorovic et al. \cite{todorovic2020arches}, from Venus to Neptune, with the Sun and Jupiter being considered the bodies $M_1$ and $M_2$ in the formulation.

First, let us define $\vec{X}=\begin{bmatrix} x & y & z & \dot{x} & \dot{y} &\dot{z} \end{bmatrix}^\text{T}$ and write Eqs. \eqref{eq:N2CRNBP} in the form:
\begin{equation}
\label{eq:XfX}
\dot{\vec{X}} = \vec{F}(\vec{X},t).
\end{equation}

The $FLI$ for the time $t$ of the system of Eqs. \eqref{eq:XfX} can be computed as \cite{froeschle2000graphical,guzzo2014evolution,lega2016theory}:
\begin{equation}
    FLI_t = \sup_{\tau \leq t } \log \left\vert \left\vert   \vec{V} (t) \right\vert  \right\vert,
\end{equation}
in which $\vec{V}(t)$ is the time evolution of a vector tangent to the initial conditions,  $\vec{V}(0)$, which can be calculated by:
\begin{equation}
    \dot{\vec{V}} (t)= \frac{\partial \vec{F}}{\partial \vec{X}} \vec{V} (t).
\end{equation}

\begin{figure}
\centering\includegraphics[width=1\textwidth]{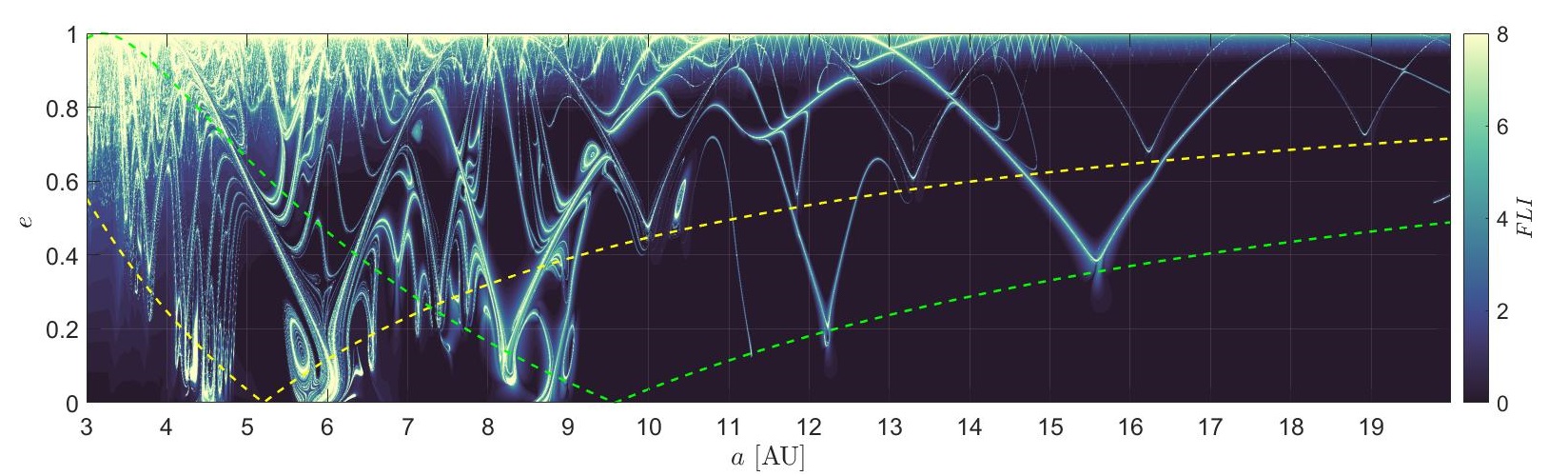}
\caption{$FLI$ applying the CRNBP.}
\label{fig:Fig3}
\end{figure}

Similar to Todorovic et al. \cite{todorovic2020arches}, it is considered a test particle of infinitesimal mass with a mean anomaly $60\degree$ ahead of Jupiter, lying on Jupiter's orbital plane. The $FLI_t$ of the test particles are computed for 100 years for different semi-major axis ($a$) and eccentricity ($e$) values relative to the sun. The computation of $FLI_t$ shows traces of the stable hyperbolic manifolds of the system because the integration will be forward in time. Figure \ref{fig:Fig3} shows the results using the CRNBP, with the ephemerides correspondence for the massive bodies being calculated accordingly to Section \ref{sec:ephemerides}. Lighter regions represent larger values of the $FLI_t$, indicating chaotic regions and traces of stable hyperbolic manifolds. The Tisserand parameter is a quasi-constant used by astronomers to distinguish between different orbits. Similar to Todorovic et al. \cite{todorovic2020arches}, a dashed yellow line depicts a value of a Tisserand parameter equals 3 in the system Sun-Jupiter, indicating the chaotic arches emanating from Jupiter close approaches. We also add a green dashed line, representing the same Tisserand parameter for Sun-Saturn and revealing the chaotic arches attached to Saturn. For a dedicated discussion and analysis of Figure \ref{fig:Fig3} we refer the reader to the original source \cite{todorovic2020arches}. Of our interest here is that the reproduction of the chaos arcs obtained by Todorovic et al. \cite{todorovic2020arches} is practically perfect. In a simple visual comparison \footnote{One can check the upper panel of the figure 1 in Todorovic et al. \cite{todorovic2020arches}.}, it appears that all structures are preserved, and none is added. The differences are minor tweaks of some of the structures for slightly, almost imperceptible, different eccentricity and semi-major axis values. 

The reproduction of the qualitatively same results of Todorovic et al. \cite{todorovic2020arches} using the CRNBP indicates that the CRNBP can be a valuable tool for exploring an ample design space before committing to a dedicated ephemerides analysis. This indication implies a much faster and simpler preliminary design, with the need to integrate only 6 ODEs instead of $6N$ ODEs of an ephemerides model. Also, the degrees of freedom of the massive bodies are reduced to a single angle $\psi_j$, allowing a straightforward analysis of multiple configurations for different epochs. Moreover, it allows a straightforward interface with a CR3BP, which we will explore soon. These advantages also contribute to the use of dynamical systems theory for identifying general dynamical behaviors, which can be explored in a multiple-body mission. The design of trajectories in the outer planetary system is an apparent advantageous application. That procedure can be much easier than patching circular restricted three-body problems, as is usually the case in the literature \cite{anderson2021tour}.

Consider Eqs. \eqref{eq:N2CRNBP} with a factor $\varepsilon$ multiplying the summations in $j$:

\begin{subequations}
\label{eq:N2CRNBP_2}
\begin{align}
\begin{split}
\ddot{x} &= 2 \dot{y} + x - \frac{\mu_1}{r_1^3}(x+\mu_2) - \frac{\mu_2}{r_2^3} (x-\mu_1) - \varepsilon \sum_{j=3}^{N-1} \mu_j \left[ \frac{1}{r_j^3} (x+\mu_2-R_j \cos \psi_j ) + \right. \\ & \left.  \sum_{k=1,k\neq j}^{N-1} \frac{\mu_k}{(R_k^2+R_j^2-2 R_k R_j \cos(\psi_k-\psi_j) )^{3/2}} (R_j \cos \psi_j - R_k \cos \psi_k ) \right], 
\end{split} \\
\begin{split}
\ddot{y} &= - 2 \dot{x} + y - \frac{\mu_1}{r_1^3} y - \frac{\mu_2}{r_2^3} y - \varepsilon \sum_{j=3}^{N-1} \mu_j \left[ \frac{1}{r_j^3} (y-R_j \sin \psi_j ) + \right. \\ & \left.  \sum_{k=1,k\neq j}^{N-1} \frac{\mu_k}{(R_k^2+R_j^2-2 R_k R_j \cos(\psi_k-\psi_j) )^{3/2}} (R_j \sin \psi_j - R_k \sin \psi_k ) \right], 
\end{split} \\
\ddot{z} &= - \frac{\mu_1}{r_1^3} z - \frac{\mu_2}{r_2^3} z - \varepsilon \sum_{j=3}^{N-1}  \frac{\mu_j}{r_j^3} z. 
\end{align}
\end{subequations}
If $\varepsilon=0$, Eqs. \eqref{eq:N2CRNBP_2} reduce to the CR3BP. Therefore, using $\varepsilon$ as a continuation parameter, dynamical features obtained in a CR3BP can be continued to the CRNBP (i.e., $\varepsilon=1$).

Let us exemplify such an approach. Using Newton's method and a continuation algorithm by pseudo-arclength, we compute a family of vertical Lyapunov orbits continued for different periods in a CR3BP, considering as the massive bodies Jupiter and Ganymede. The family of computed vertical Lyapunov orbits about the Lagrangean point $\mathcal{L}_3$ are shown in Figure \ref{fig:Fig4}. We also formulate a CRNBP for the bodies Jupiter, Ganymede, Europa, and Io, considering an ephemerides correspondence to April 9, 2016, 00:00:00 TDB, and making the period of Io and Europe exactly a quarter and a half of Ganymede's mean orbital period, respectively. Therefore, the CRNBP for that system is a periodic non-autonomous system, such that:

\begin{equation}
\label{eq:XfX}
\dot{\vec{X}} = \vec{F}(\vec{X},t) = \vec{F}(\vec{X},t+pT),
\end{equation}
in which $p\in \mathbb{Z}$, and $T$ is the period of the nonlinear system of equations, which is the mean orbital period of Ganymede for that CRNBP composed of Jupiter-Ganymede-Io-Europa. Thus, such a system of equations can only possess a periodic orbit if its period is equal to $pT$.

We choose the vertical Lyapunov orbit from the family shown in Figure \ref{fig:Fig4} with a period equal to the Ganymede's period ($p=1$) and then continue it to the CRNBP through the parameter $\varepsilon$. In Figure \ref{fig:Fig5} is shown the continued orbits, from $\varepsilon=0$ (CR3BP) to $\varepsilon=1$ (CRNBP). The lightest blue color indicates the periodic orbit in the CR3BP, with its projection on each plane depicted by the lightest orange color. The pinkest color indicates the periodic orbit in the CRNBP, and its projection on each plane is depicted in black. As one can note, the deformation in size of the periodic orbit when continuing it to the CRNBP is considerably large, with magnitudes in the order of half of the Ganymede's semi-major axis (1 c.u.). Similarly, periodic orbits obtained in the CR3BP can be continued to quasi-periodic torii in the CRNBP, which can also be useful if including Callisto in the system, as made by McCarthy and Howell \cite{mccarthy2021quasi} using the BCR4BP for Earth-Moon-Sun.

\begin{figure}
\centering\includegraphics[width=.5\textwidth]{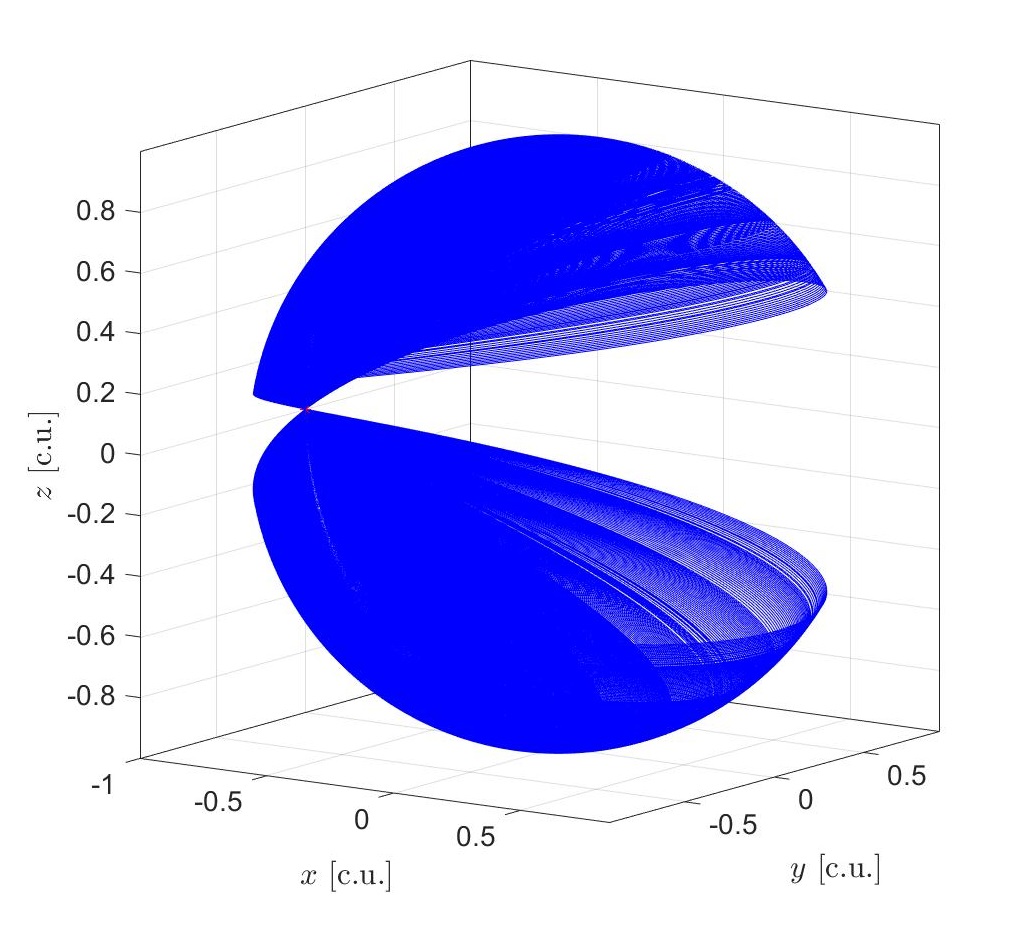}
\caption{Lyapunov vertical orbit family about $\mathcal{L}_3$, using the CR3BP for Jupiter-Ganymede.}
\label{fig:Fig4}
\end{figure}

\begin{figure}
\centering\includegraphics[width=.5\textwidth]{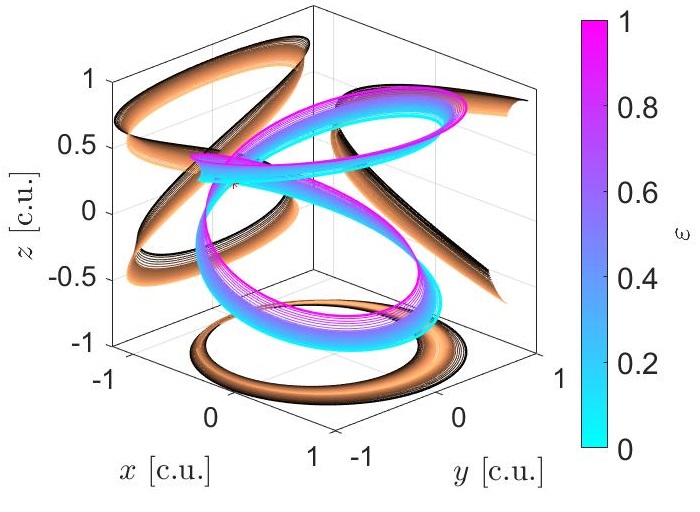}
\caption{Continuation of a Lyapunov vertical orbit about $\mathcal{L}_3$, using the CR3BP for Jupiter-Ganymede, to a periodic orbit in the CRNBP considering the Laplace resonance of Ganymede-Europa-Io.}
\label{fig:Fig5}
\end{figure}

Some authors have proposed the use of periodic orbits and associated hyperbolic manifolds to design low-energy trajectories for landing in outer planetary systems' moons \cite{bury2020landing,hernandez2020landing}. In the next example we follow that trend and try to show the rich behavior of trajectories when considering a CRNBP, which can pass unnoticed if considering only the CR3BP. To avoid deviating from this work's scope, and letting a more complex analysis for dedicated papers, here suffices to show the richness of trajectories straightforwardly. For that sake, we consider a CR3BP for the system Jupiter-Europa and a CRNBP for Jupiter-Europa-Io-Ganymede-Callisto\footnote{Europa is chosen as the body $M_2$, Figure \ref{fig:Fig2}.}. The initial position for the spacecraft, $\vec{r}_0=\begin{bmatrix} x_0 & y_0 & z_0 \end{bmatrix}^\text{T}$, is set 50 km above Europa's surface, for different longitude angles $\theta$ measured from the $x$ axis in Figure \ref{fig:Fig2}, with the spacecraft restrained to lie on the orbital plane of the massive bodies ($z_0=\dot{z}_0=0$). The initial velocity, $\vec{v}_0=\begin{bmatrix} \dot{x}_0 & \dot{y}_0 & 0 \end{bmatrix}^\text{T}$, is calculated from the Jacobi constant of the Lagrangean point $\mathcal{L}_2$, $J_{\mathcal{L}_2}$, of the CR3BP by:
\begin{equation}
    \vert \vert \vec{v}_0 \vert \vert ^2 = x_0^2 + y_0^2 + 2 \left( \frac{\mu_1}{r_1} + \frac{\mu_2}{r_2} \right) - J_{\mathcal{L}_2},
\end{equation}
 with the velocity direction chosen for a prograde motion with respect to Europa and perpendicular to the spacecraft's initial position, i.e.:
 \begin{equation}
     \vec{v}_0 = \vert \vert \vec{v}_0 \vert \vert \frac{\hat{z} \times \vec{r}_0}{\vert \vert \hat{z} \times \vec{r}_0 \vert \vert },
 \end{equation}
where $\hat{z}$ is the unit vector in the direction of the z axis. The ephemerides corresponde for the initial state is chosen as April 9, 2016, 00:00:00 TDB. 

\begin{figure}[H]
\centering
\subfloat[System view]{\includegraphics[width=.5\textwidth]{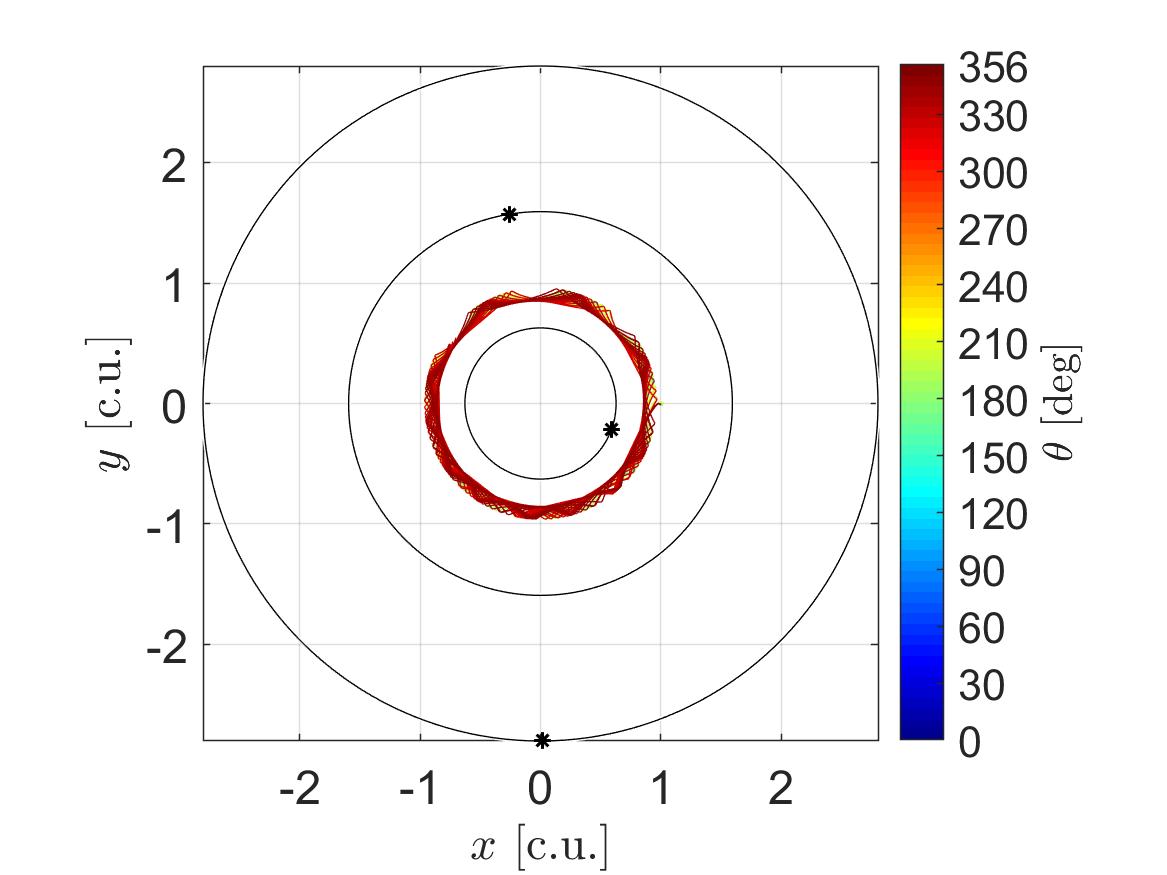}\label{Fig6a}}
\subfloat[Europa close up view]{\includegraphics[width=.5\textwidth]{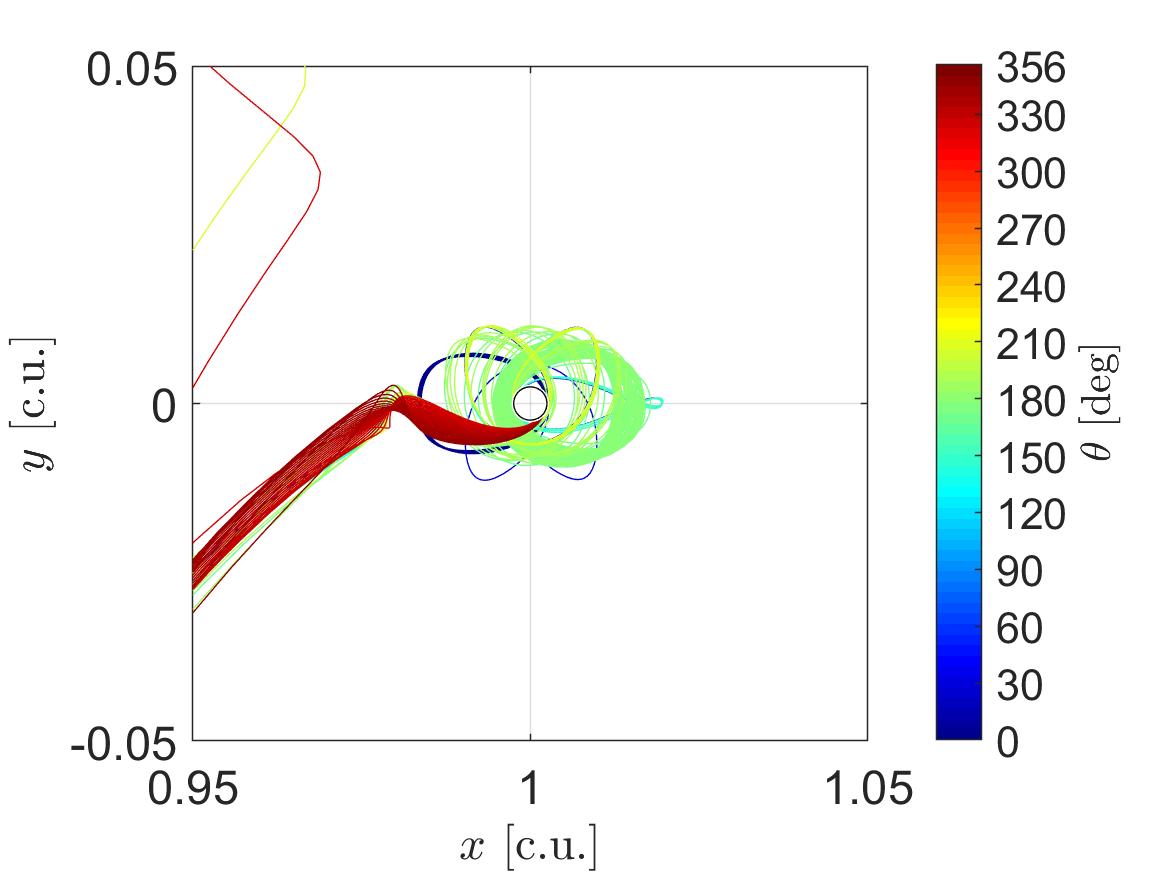}\label{Fig6b}} 

\caption{Trajectories integrated backward in time using the CR3BP, Jupiter-Europa.}
\label{Fig6}
\end{figure}

\begin{figure}[H]
\centering
\subfloat[$t_a=0$ days]{\includegraphics[width=.5\textwidth]{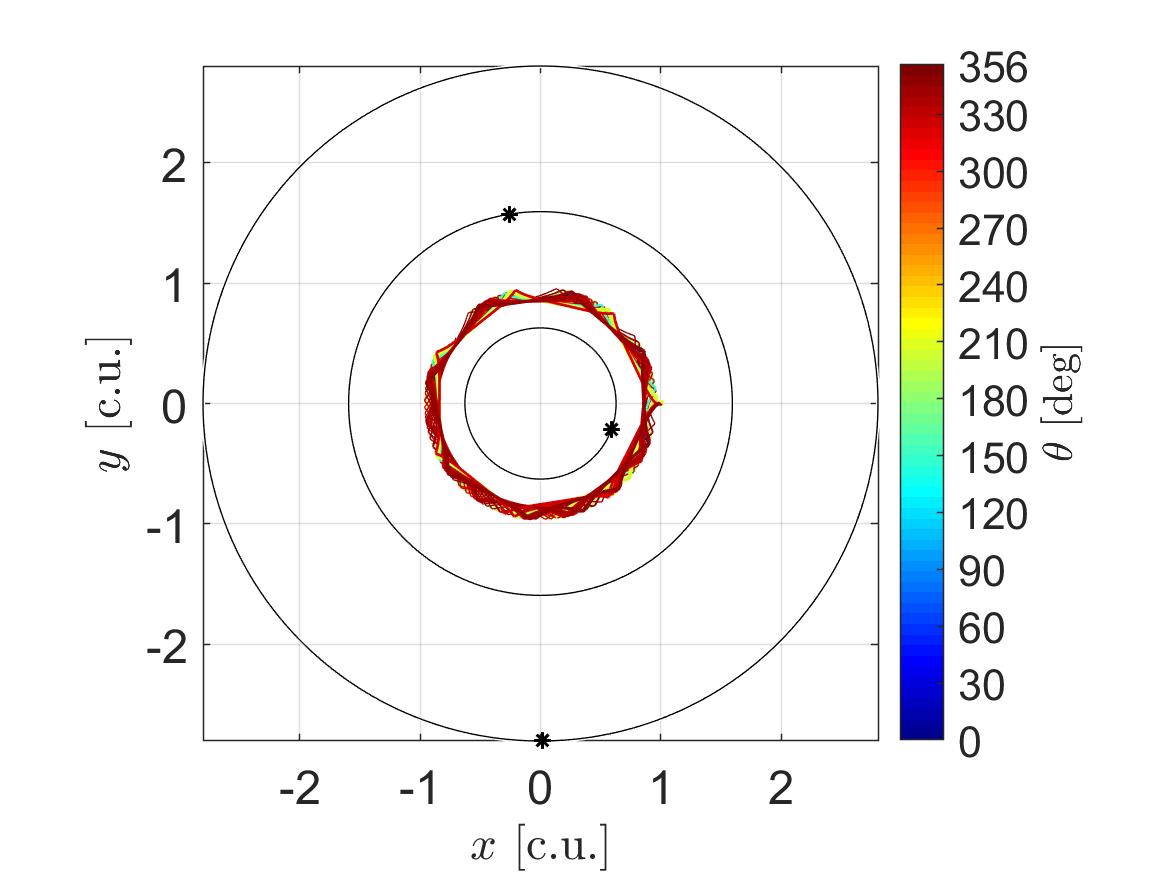}\label{Fig7a}}
\subfloat[$t_a=2.36$ days]{\includegraphics[width=.5\textwidth]{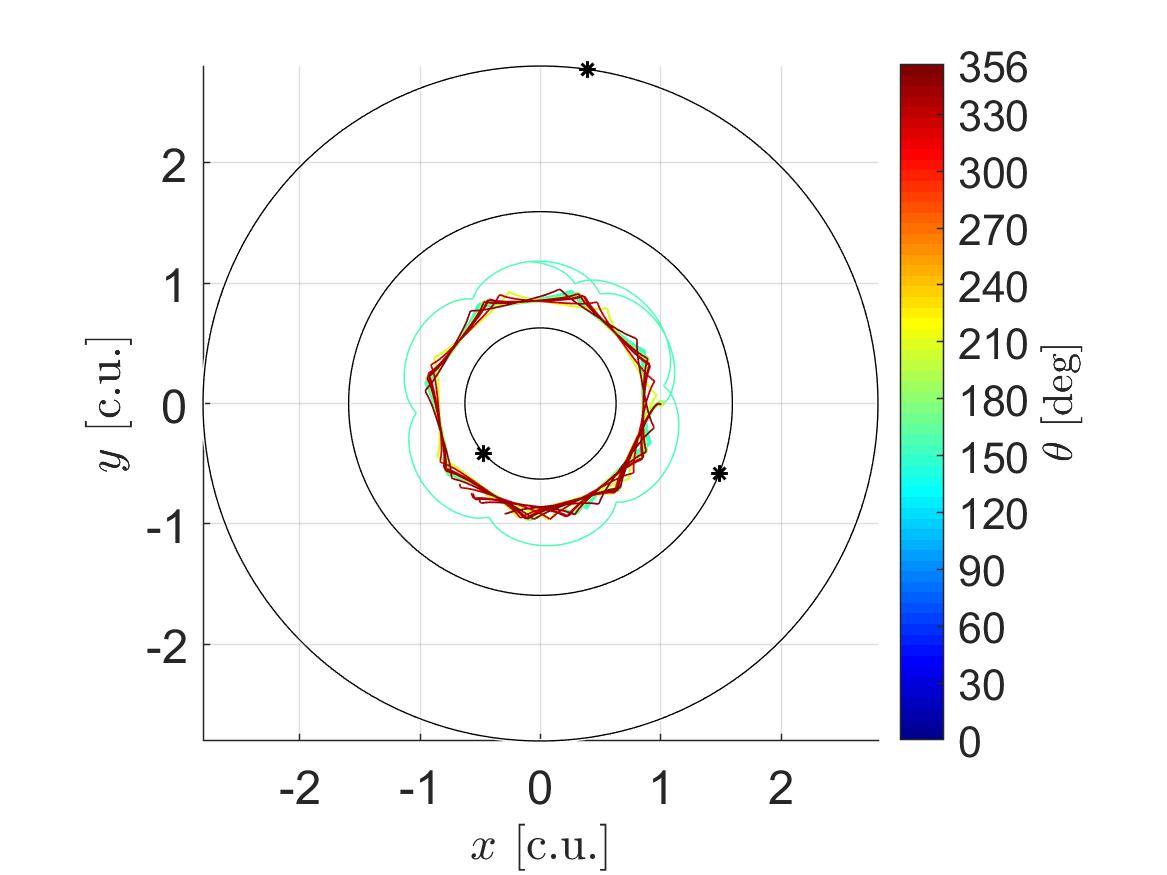}\label{Fig7c}}  \\
\subfloat[$t_a=7.1$ days]{\includegraphics[width=.5\textwidth]{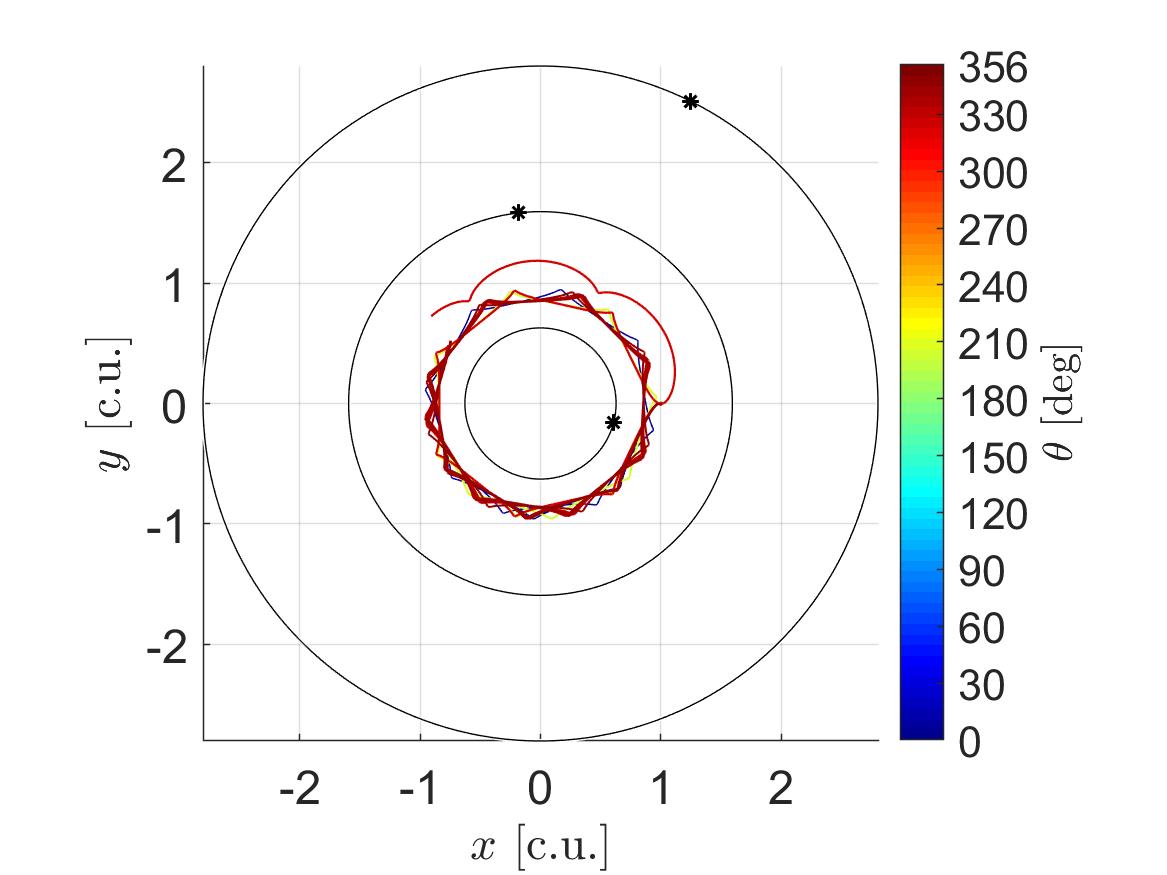}\label{Fig7f}}

\caption{Trajectories integrated backward in time using the CRNBP, Jupiter-Europa-Io-Ganymede-Callisto for different arrival times, April 9, 2016, 00:00:00 $+ t_a$ days.}
\label{Fig7}
\end{figure}

\begin{figure}[H]
\centering
\subfloat[$t_a=0$ days]{\includegraphics[width=.5\textwidth]{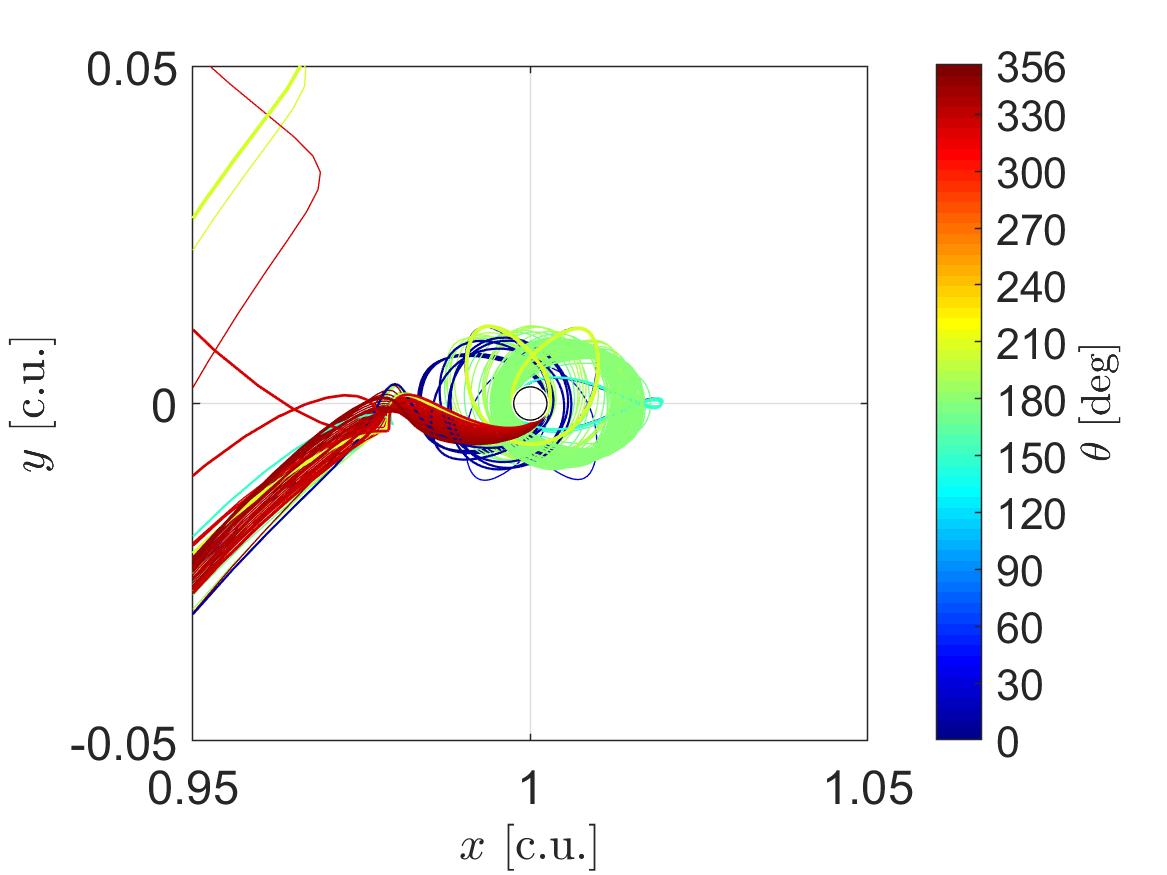}\label{Fig8a}}
\subfloat[$t_a=2.36$ days]{\includegraphics[width=.5\textwidth]{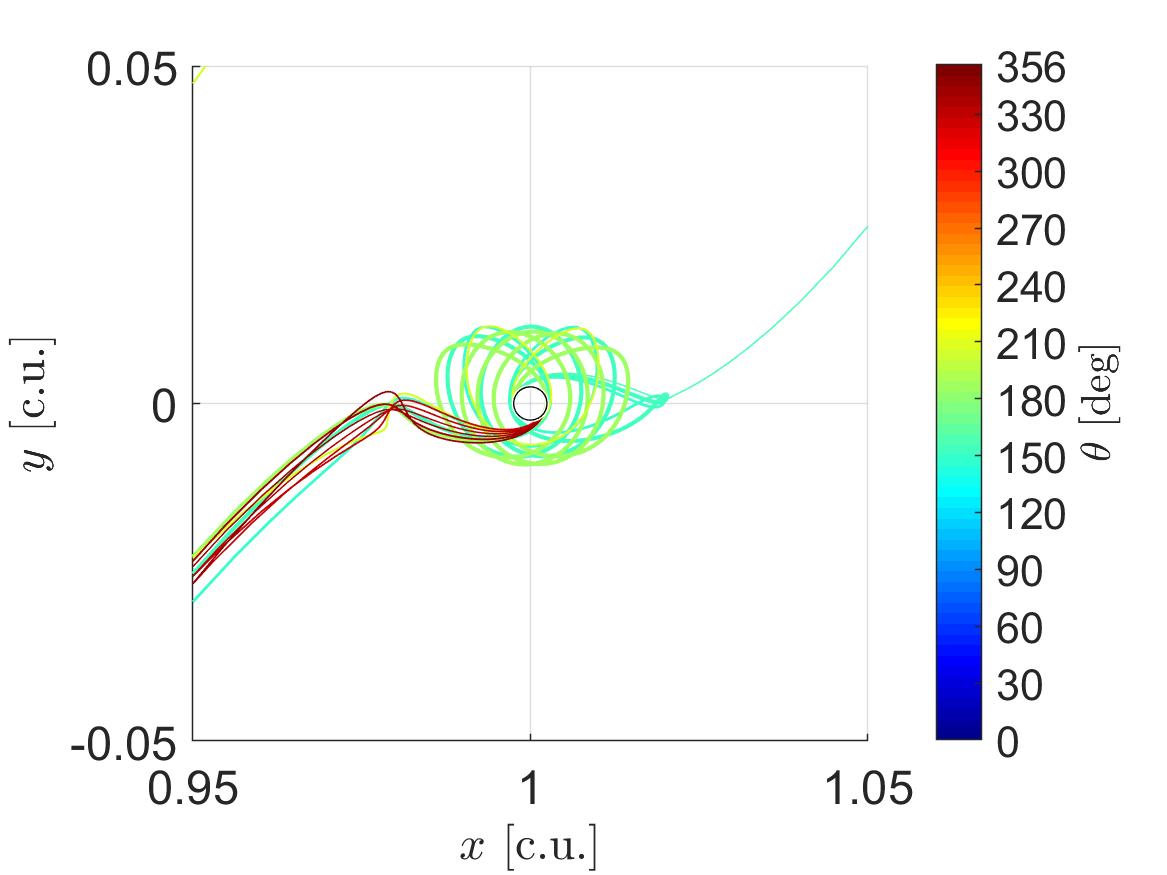}\label{Fig8c}}  \\
\subfloat[$t_a=7.1$ days]{\includegraphics[width=.5\textwidth]{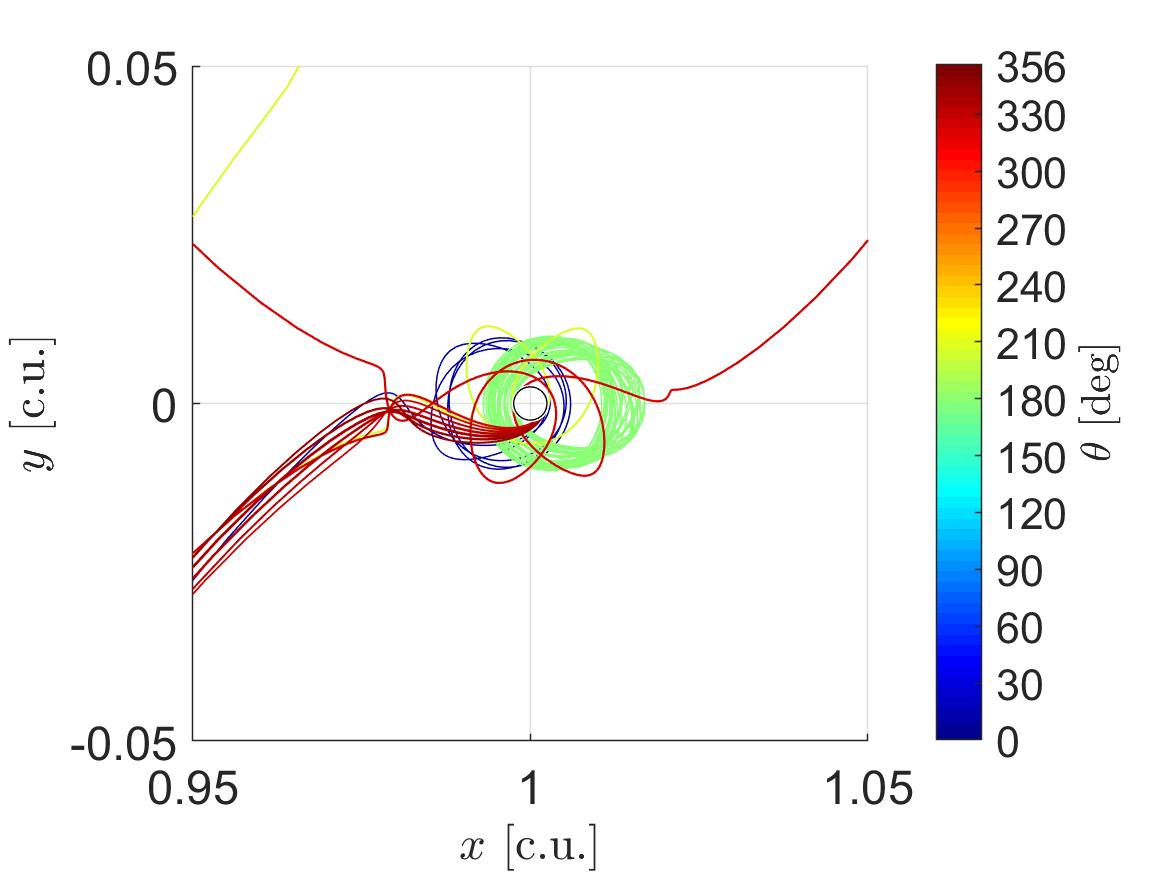}\label{Fig8f}}

\caption{Trajectories integrated backward in time using the CRNBP, Jupiter-Europa-Io-Ganymede-Callisto for different arrival times, April 9, 2016, 00:00:00 $+ t_a$ days.}
\label{Fig8}
\end{figure}

With the chosen initial state, the equations of motion are integrated backward in time up to 30 days before April 9, 2016, 00:00:00 TDB. The trajectories obtained with the CR3BP are depicted in Figure \ref{Fig6}, where Figure \ref{Fig6a} shows the whole system point of view\footnote{Although the other moons are represented in the figure, they are not taken into account in the CR3BP.}, and Figure \ref{Fig6b} shows the trajectories nearby Europa. Trajectories that collide Europa are omitted in the plots. In the same manner, Figures \ref{Fig7} and \ref{Fig8} show the trajectories in the system perspective and close to Europa, respectively, and using the CRNBP for different arrival times $t_a$ after April 9, 2016, 00:00:00 TDB. The differences in considering the other bodies are quite obvious\footnote{A video for different arrival times can be accessed in \url{https://www.youtube.com/watch?v=5rXJ3DFNnJw&list=PLL7PeNhFNiitH3rOpTBx6Tf9UeRGCqFT8}, which makes much more clear the richness of trajectories in the CRNBP.}. A distinguishing feature is the opening for trajectories beyond Europa's orbit, just changing the arrival epoch. Figures \ref{Fig7c} and \ref{Fig8c} show the possibility of arriving from a transeuropa region for a landing with an angle $\theta$ around $150\degree$ when delaying the arrival 2.36 days after April 9, 2016, 00:00:00 TDB. For a $\theta$ around $350\degree$, there is a transeuropa trajectory when delaying the arrival to 7.1 days after April 9, 2016, 00:00:00 TDB, as shown in Figures \ref{Fig7f} and \ref{Fig8f}. The CRNBP provides easiness in exploring such features. For instance, similarly to other works using the CR3BP \cite{bury2020landing,hernandez2020landing}, the equilibrium point $\mathcal{L}_2$ of the CR3BP can be continued to a quasi-periodic torus in that CRNBP, and hyperbolic stable manifolds bringing to the torus can be calculated and patched to unstable hyperbolic manifolds passing close to Europa \cite{olikara2012numerical}.

\section{Conclusions}

In this brief Note, we derive the equations of motion for what we are calling a circular restricted n-body problem, CRNBP. We can achieve this by using the same reasoning used in the famous bicircular restricted four-body problem, BCR4BP, which is imposing artificial dynamical constraints in the movement of the massive bodies, i.e., assuming that all the massive bodies can only describe a Keplerian motion. A much faster and simpler preliminary design is possible with the CRNBP, needing to integrate only six first-order ordinary differential equations instead of the 6N of an ephemerides model. We reproduce a complex dynamical behavior observed in an ephemerides n-body problem, indicating the structural stability of the CRNBP. Next, we show some advantages in applying the CRNBP, especially for designing trajectories in outer planetary systems. The application of dynamical systems theory is of particular interest due to the simplicity of the CRNBP and its friendly interface with a CR3BP, making it more straightforward to extend results from the latter to an n-body problem. We show that a multiple body system hides many subtleties that are challenging to uncover with a CR3BP only analysis.

\section*{Funding Sources}

The authors wish to express their appreciation for the support provided by grant $\#$ 309089/2021-2 from the  National Council for Scientific and Technological Development (CNPq); grants $\#$ 2017/20794-2 and 2016/24561-0 from S\~ao Paulo Research Foundation (FAPESP); and the financial support from the Coordination for the Improvement of Higher Education Personnel (CAPES). The publication has been prepared with the support of the RUDN University Strategic Academic Leadership Program.  


\bibliographystyle{ieeetr}
\bibliography{referencia}

\begin{thebibliography}{10}

\bibitem{tsander1964}
F.~A. Tsander, {\em Problems of Flight by Jet Propulsion}, ch.~Flights to Other
  Planets (The Theory of Interplanetary Travel).
\newblock Israel Program for Scientific Translations, 2nd~ed., 1964.
\newblock English translation of Problema poleta pri pomoshehi reaktivnykh
  apparatov.

\bibitem{negri2020historical}
R.~B. Negri and A.~F.~B. de~Almeida~Prado, ``A historical review of the theory
  of gravity-assists in the pre-spaceflight era,'' {\em Journal of the
  Brazilian Society of Mechanical Sciences and Engineering}, vol.~42, no.~8,
  pp.~1--10, 2020.

\bibitem{lawden1954perturbation}
D.~Lawden, ``Perturbation maneuvers,'' {\em Journal of the British
  Interplanetary Society}, vol.~13, no.~5, 1954.

\bibitem{crocco1956one}
G.~A. Crocco, ``One-year exploration-trip earth-mars-venus-earth,'' {\em
  Proceedings of the VIIth International Astronautical Congress Rome},
  pp.~227--252, 1956.

\bibitem{ehricke1957instrumented}
K.~A. Ehricke, ``Instrumented comets—astronautics of solar and planetary
  probes,'' in {\em Proceedings of the 8th International Astronautical
  Congress, Barcelona}, no.~493-57, pp.~74--126, American Rocket Society, 1957.

\bibitem{minovitch1961method}
M.~Minovitch, ``A method for determining interplanetary free-fall
  reconnaissance trajectories,'' {\em JPL Tec. Memo}, vol.~312, p.~130, 1961.

\bibitem{negri2017studying}
R.~B. Negri, A.~F.~B. de~Almeida~Prado, and A.~Sukhanov, ``Studying the errors
  in the estimation of the variation of energy by the “patched-conics”
  model in the three-dimensional swing-by,'' {\em Celestial Mechanics and
  Dynamical Astronomy}, vol.~129, no.~3, pp.~269--284, 2017.

\bibitem{negri2019lunar}
R.~B. Negri, A.~Sukhanov, and A.~F.~B. de~Almeida~Prado, ``Lunar gravity
  assists using patched-conics approximation, three and four body problems,''
  {\em Advances in Space Research}, vol.~64, no.~1, pp.~42--63, 2019.

\bibitem{szebehely2012theory}
V.~SZEBEHELY, ``Chapter 1 - description of the restricted problem,'' in {\em
  Theory of Orbit} (V.~SZEBEHELY, ed.), pp.~7--41, Academic Press, 1967.

\bibitem{egorov1958certain}
V.~A. Egorov, ``Certain problems of moon flight dynamics,'' {\em The Russian
  Literature of Satellites}, pp.~107--174, 1958.
\newblock English translation of O nekotorykh zadachakh dinamiki poleta k Lune.

\bibitem{huang1960very}
S.-S. Huang, ``Very restricted four-body problem.,'' {\em The Astronomical
  Journal}, vol.~70, p.~347, 1960.

\bibitem{negri2020generalizing}
R.~B. Negri and A.~F. Prado, ``Generalizing the bicircular restricted four-body
  problem,'' {\em Journal of Guidance, Control, and Dynamics}, vol.~43, no.~6,
  pp.~1173--1179, 2020.

\bibitem{howell1986periodic}
K.~C. Howell and D.~B. Spencer, ``Periodic orbits in the restricted four-body
  problem,'' {\em Acta Astronautica}, vol.~13, no.~8, pp.~473--479, 1986.

\bibitem{belbruno1993sun}
E.~A. Belbruno and J.~K. Miller, ``Sun-perturbed earth-to-moon transfers with
  ballistic capture,'' {\em Journal of Guidance, Control, and Dynamics},
  vol.~16, no.~4, pp.~770--775, 1993.

\bibitem{castella2000vertical}
E.~Castella and {\`A}.~Jorba, ``On the vertical families of two-dimensional
  tori near the triangular points of the bicircular problem,'' {\em Celestial
  Mechanics and Dynamical Astronomy}, vol.~76, no.~1, pp.~35--54, 2000.

\bibitem{mingotti2011optimal}
G.~Mingotti, F.~Topputo, and F.~Bernelli-Zazzera, ``Optimal low-thrust
  invariant manifold trajectories via attainable sets,'' {\em Journal of
  guidance, control, and dynamics}, vol.~34, no.~6, pp.~1644--1656, 2011.

\bibitem{qi2014gravitational}
Y.~Qi, S.~Xu, and R.~Qi, ``Gravitational lunar capture based on bicircular
  model in restricted four body problem,'' {\em Celestial Mechanics and
  Dynamical Astronomy}, vol.~120, no.~1, pp.~1--17, 2014.

\bibitem{heiligers2018solar}
J.~Heiligers and D.~J. Scheeres, ``Solar-sail orbital motion about asteroids
  and binary asteroid systems,'' {\em Journal of Guidance, Control, and
  Dynamics}, vol.~41, no.~9, pp.~1947--1962, 2018.

\bibitem{boudad2020dynamics}
K.~K. Boudad, K.~C. Howell, and D.~C. Davis, ``Dynamics of synodic resonant
  near rectilinear halo orbits in the bicircular four-body problem,'' {\em
  Advances in Space Research}, vol.~66, no.~9, pp.~2194--2214, 2020.

\bibitem{gabern2001restricted}
F.~Gabern and A.~Jorba, ``A restricted four-body model for the dynamics near
  the lagrangian points of the sun-jupiter system,'' {\em Discrete \&
  Continuous Dynamical Systems-B}, vol.~1, no.~2, p.~143, 2001.

\bibitem{pergola2009earth}
P.~Pergola, K.~Geurts, C.~e.~a. Casaregola, and M.~Andrenucci, ``Earth--mars
  halo to halo low thrust manifold transfers,'' {\em Celestial Mechanics and
  Dynamical Astronomy}, vol.~105, no.~1, pp.~19--32, 2009.

\bibitem{oshima2015jumping}
K.~Oshima and T.~Yanao, ``Jumping mechanisms of trojan asteroids in the planar
  restricted three-and four-body problems,'' {\em Celestial Mechanics and
  Dynamical Astronomy}, vol.~122, no.~1, pp.~53--74, 2015.

\bibitem{barrabes2016pseudo}
E.~Barrab{\'e}s, G.~G{\'o}mez, J.~M. Mondelo, and M.~Oll{\`e},
  ``Pseudo-heteroclinic connections between bicircular restricted four-body
  problems,'' {\em Monthly Notices of the Royal Astronomical Society},
  vol.~462, no.~1, pp.~740--750, 2016.

\bibitem{iuliano2016solution}
J.~R. Iuliano, ``A solution to the circular restricted n body problem in
  planetary systems,'' 2016.

\bibitem{wiggins1990introduction}
S.~Wiggins, ``Structural stability, genericity, and transversality,'' in {\em
  Introduction to Applied Nonlinear Dynamical Systems and Chaos}, pp.~157--168,
  New York, NY: Springer New York, 2003.

\bibitem{todorovic2020arches}
N.~Todorovi{\'c}, D.~Wu, and A.~J. Rosengren, ``The arches of chaos in the
  solar system,'' {\em Science advances}, vol.~6, no.~48, p.~eabd1313, 2020.

\bibitem{froeschle1997fast}
C.~Froeschl{\'e}, E.~Lega, and R.~Gonczi, ``Fast lyapunov indicators.
  application to asteroidal motion,'' {\em Celestial Mechanics and Dynamical
  Astronomy}, vol.~67, no.~1, pp.~41--62, 1997.

\bibitem{froeschle2000graphical}
C.~Froeschl{\'e}, M.~Guzzo, and E.~Lega, ``Graphical evolution of the arnold
  web: from order to chaos,'' {\em Science}, vol.~289, no.~5487,
  pp.~2108--2110, 2000.

\bibitem{guzzo2014evolution}
M.~Guzzo and E.~Lega, ``Evolution of the tangent vectors and localization of
  the stable and unstable manifolds of hyperbolic orbits by fast lyapunov
  indicators,'' {\em SIAM Journal on Applied Mathematics}, vol.~74, no.~4,
  pp.~1058--1086, 2014.

\bibitem{lega2016theory}
E.~Lega, M.~Guzzo, and C.~Froeschl{\'e}, ``Theory and applications of the fast
  lyapunov indicator (fli) method,'' in {\em Chaos Detection and
  Predictability}, pp.~35--54, Springer, 2016.

\bibitem{anderson2021tour}
R.~L. Anderson, ``Tour design using resonant-orbit invariant manifolds in
  patched circular restricted three-body problems,'' {\em Journal of Guidance,
  Control, and Dynamics}, vol.~44, no.~1, pp.~106--119, 2021.

\bibitem{mccarthy2021quasi}
B.~McCarthy and K.~Howell, ``Quasi-periodic orbits in the sun-earth-moon
  bicircular restricted four-body problem,'' in {\em Proceedings of the 31st
  AAS/AIAA Space Flight Mechanics Meeting}, AAS/AIAA, 2021.

\bibitem{bury2020landing}
L.~Bury and J.~W. McMahon, ``Landing trajectories to moons from the unstable
  invariant manifolds of periodic libration point orbits,'' in {\em AIAA
  Scitech 2020 Forum}, p.~2181, 2020.

\bibitem{hernandez2020landing}
S.~Hernandez, R.~Anderson, D.~Roth, Y.~Takahashi, and T.~McElrath, ``Navigating
  low-energy trajectories to land on the surface of europa,'' in {\em
  Proceedings of the 31st AAS/AIAA Space Flight Mechanics Meeting}, AAS/AIAA,
  2021.

\bibitem{olikara2012numerical}
Z.~P. Olikara and D.~J. Scheeres, ``Numerical method for computing
  quasi-periodic orbits and their stability in the restricted three-body
  problem,'' {\em Advances in the Astronautical Sciences}, vol.~145,
  no.~911-930, 2012.

\end{thebibliography}

\end{document}